\documentclass[12ppt,twocolumns]{emulateapj}
\usepackage{rotfloat}

\begin{document}
\shorttitle{TURBULENT RECONNECTION ON PROTOSTELLAR DISKS}
\shortauthors{SANTOS-LIMA, DE GOUVEIA DAL PINO \& LAZARIAN}
\title{The role of turbulent magnetic reconnection  on the formation of rotationally supported protostellar disks}

\author{R. Santos-Lima\altaffilmark{1}, E. M. de Gouveia Dal Pino\altaffilmark{1},  A. Lazarian\altaffilmark{2}}
\altaffiltext{1}{Instituto de Astronomia, Geof\'isica e Ci\^encias Atmosf\'ericas, Universidade de S\~ao Paulo, R. do Mat\~ao, 1226, S\~ao Paulo, SP 05508-090, Brazil}
\altaffiltext{2}{Department of Astronomy, University of Wisconsin, Madison, WI 53706, USA}

\begin{abstract}
The  formation of protostellar disks out of molecular cloud cores
is still not fully understood. Under ideal MHD
conditions, the removal of angular momentum from the  disk progenitor by the typically
embedded  magnetic field may prevent the formation of a rotationally supported disk
during the main protostellar accretion phase of low mass stars. This has been known as the
magnetic braking problem and the most investigated mechanism to
alleviate this problem and help removing the excess of magnetic flux
during the star formation process, the so called ambipolar diffusion
(AD),  has been shown  to be not sufficient to weaken the magnetic
braking at least at this stage of the disk formation.  In this work, motivated by recent progress
in the understanding of magnetic reconnection in turbulent environments, we appeal to
the diffusion of magnetic field mediated by magnetic reconnection as an alternative mechanism
for removing magnetic flux. We investigate numerically this
mechanism during the  later phases of the protostellar disk
formation and show its high efficiency. By means of  fully 3D MHD simulations, we show that
the diffusivity arising from turbulent magnetic reconnection is able to transport
magnetic flux to the outskirts of the disk progenitor at time scales
compatible with the collapse, allowing the formation of a
rotationally supported  disk  around the protostar of dimensions $\sim$ 100 AU, with a nearly
Keplerian profile in the early accretion phase. Since MHD turbulence is expected to be present in protostellar disks,
this is a  natural mechanism  for removing
magnetic flux excess and allowing the formation of these disks. This mechanism dismiss the necessity of postulating a hypothetical
increase of the Ohmic resistivity as discussed in the literature.
Together with our earlier work which showed that magnetic flux removal  from
molecular cloud cores is very efficient, this work calls for reconsidering the relative role of AD
for the processes of star and planet formation.

\end{abstract}


\section{Introduction}

Circumstellar disks (with typical masses $\sim 0.1$ M$_{\odot}$ and
diameters $\sim 100$ AU) are known to play a fundamental role in the
late stages of star formation and also in planet formation. However,
the mechanism that allows their formation and the decoupling from
the surrounding molecular cloud core progenitor is still not fully
understood (see, e.g., \citealt{krasnopolsky_etal_2011} for a  recent comprehensive
review). Former studies have shown that the observed embedded
magnetic fields in molecular cloud cores, which imply magnetic mass-to-flux ratios relative to the critical value a few times larger than unity (\citealt{crutcher_2005, troland_crutcher_2008}) are high enough to inhibit the
formation of rationally supported disks during the main protostellar
accretion phase of low mass stars, provided that ideal MHD applies.
This has been known as the magnetic braking problem (see e.g.,
\citealt{allen_etal_2003, galli_etal_2006, price_bate_2007, hennebelle_fromang_2008, mellon_li_2008}).

Proposed mechanisms to alleviate this problem and help removing the
excess of magnetic flux during the star formation process  include
non-ideal MHD effects such
as  ambipolar diffusion (AD) and, to a smaller  degree, Ohmic dissipation effects.
The AD, which was first discussed in this context
by \citet{mestel_spitzer_1956}, has been extensively  investigated since
then (e.g., \citealt{spitzer_1968, nakano_tademaru_1972, mouschovias_1976,
mouschovias_1977, mouschovias_1979, nakano_nakamura_1978, shu_1983, lizano_shu_1989,
fiedler_mouschovias_1992, fiedler_mouschovias_1993, li_etal_2008, fatuzzo_adams_2002, zweibel_2002}). In principle, AD  allows magnetic flux to be
redistributed during the collapse in low ionization regions as the
result of the differential motion between the ionized and the neutral
gas.   However, for  realistic  levels  of  core  magnetization  and
ionization,  recent work has shown that AD does not seem to be
sufficient to weaken the magnetic braking in order to allow rotationally
supported disks  to  form.  In  some cases, the magnetic braking has
been found to be even enhanced by AD  \citep{mellon_li_2009,
krasnopolsky_konigl_2002, basu_mouschovias_1995, hosking_whitworth_2004, duffin_pudritz_2009, li_etal_2011}.
\footnote{See however a recent work that investigates the effects of  AD in the  triggering of magneto-rotational instability in more evolved  cold, proto-planetary disks where the  fraction of neutral gas is much larger \citep{bai_stone_2011}.} These  findings motivated
\citet{krasnopolsky_etal_2010} (see also \citealt{li_etal_2011})  to  examine whether Ohmic
dissipation could be effective in  weakening the magnetic  braking.
They claimed that in order to  enable the formation of persistent,
rotationally supported disks during the protostellar mass accretion
phase a highly enhanced resistivity, or ``hyper-resistivity''
$\eta \gtrsim 10^{19}$ cm$^2$s$^{-1}$ of unspecified origin would be required. Although this value  is somewhat dependent
on the degree of core magnetization, it implies that the
required resistivity is a few orders of magnitude larger than the
classic microscopic Ohmic resistivity values \citep{krasnopolsky_etal_2010}.

On the other hand, \citet{machida_etal_2010} (see also \citealt{inutsuka_etal_2010, machida_etal_2011}) performed core collapse three-dimensional simulations and found
that, even with just the classical Ohmic resistivity, a tinny  rotationally supported disk can form at
the beginning of the protostellar accretion phase (see also \citealt{dapp_basu_2010}) and grow to
larger, 100-AU scales at later times.  They claim that the later growth of
the circumstellar disk is caused by the depletion of the infalling envelope.  As long as  this envelope remains more massive than the circumstellar disk, the 
magnetic braking is effective, but  when the circumstellar disk becomes more massive, then the envelope cannot brake the disk anymore. In their simulations, they assume an initially much denser core than in Krasnopolsky et al.  work, which helps the early formation of a tiny rotating disk facilitated by the Ohmic diffusion in the central regions. But they have to wait for over $10^5$ yr  in order to allow a large-scale rotationally supported, massive disk to form (see discussion in \S5).  While this question on the effectiveness  of the Ohmic diffusion in the early accretion phases of disk formation deserves further careful testing,  we here discuss an alternative more efficient mechanism to diffuse the magnetic flux based on turbulent reconnection.


Before addressing this new mechanism, it is crucial to note first that 
the concept of ``hyper-resistivity'' previously mentioned (see also
\citealt{strauss_1986, bhattacharjee_hameiri_1986, diamond_malkov_2003}) is not physically justified
and therefore one  cannot rely on it (see criticism in \citealt{kowal_etal_2009, eyink_etal_2011}). Therefore, the dramatic increase of resistivity is not justified.

However, one may notice that the problem we face with the magnetic field is not of dissipation of magnetic flux, where, indeed, ordinary resistivity is necessary, but of magnetic diffusion. Frequently in the literature, magnetic diffusion is disregarded for highly conductive astrophysical flows. This is based on the concept of magnetic field being frozen-in into conductive fluids and thus moving together with the fluid. However, the frozen-in condition is true in the absence of magnetic reconnection. When the latter is present,  changes in the magnetic field topology are allowed.

On the basis of \citet{lazarian_vishniac_1999} (henceforth LV99) model of reconnection, \citet{lazarian_2005} (henceforth L05) suggested a way of removing magnetic flux
in the process of star formation, presenting the concept of ``reconnection diffusion'' of
magnetic field in turbulent media (see also \citealt{lazarian_vishniac_2009}). This
concept was applied successfully to star formation in \citet{santos-lima_etal_2010} (see also
discussion in \citealt{lazarian_etal_2010, de_gouveia_dalpino_etal_2011}, and new numerical calculations in \citealt{leao_etal_2011}).
There, \citet{santos-lima_etal_2010} (henceforth SX10) performed 3D MHD simulations of
turbulent flows and found a decoupling between the gas and the
magnetic field due to reconnection under one-fluid approximation,
i.e.,  without having ambipolar diffusion. In addition, in the
presence of gravity, they found a decrease of the magnetic
flux-to-mass ratio with increasing gas density in the center of the
gravitational potential well, both  for systems starting  with
equilibrium distributions of gas and magnetic field and  for
dynamically unstable, collapsing systems.  This provides evidence that the
process of magnetic flux removal  by turbulent reconnection
diffusion is important to quasi-static subcritical
clouds  and also to collapsing supercritical cores. Thus it is natural to explore the consequences of the process in other systems.

In this work, we investigate this effect on the removal of magnetic
flux from  collapsing, rotating protostellar cores. \citet{lazarian_vishniac_2009}
argued  that the removal of magnetic fields from circumstellar disks reported in the work of \citet{shu_etal_2006} is due to processes of turbulent reconnection, but have
not provided any quantitative study of the effect. A preliminary discussion of
the quantitative results can be found in \citet{de_gouveia_dalpino_etal_2011}.
Here, we show by means of 3D MHD numerical simulations that turbulent magnetic
reconnection diffusivity enables the transport of magnetic
flux to the outskirts of the core at time scales compatible with the
collapse time scale, thus  allowing the formation of a
rotationally supported protostelar disk with nearly Keplerian
profile.

In \S2, we discuss the theoretical foundations of our work, in \S3 we describe the numerical setup and initial conditions, in \S4 we show
the results of our three-dimensional (3D) MHD turbulent  numerical simulations of disk formation, and
in \S5, we  discuss our results within a bigger picture of reconnection diffusion processes. Our summary is presented in \S6.

\section{Theoretical considerations}

\begin{figure}[!t]
 \begin{center}
\includegraphics[width=1.0 \columnwidth]{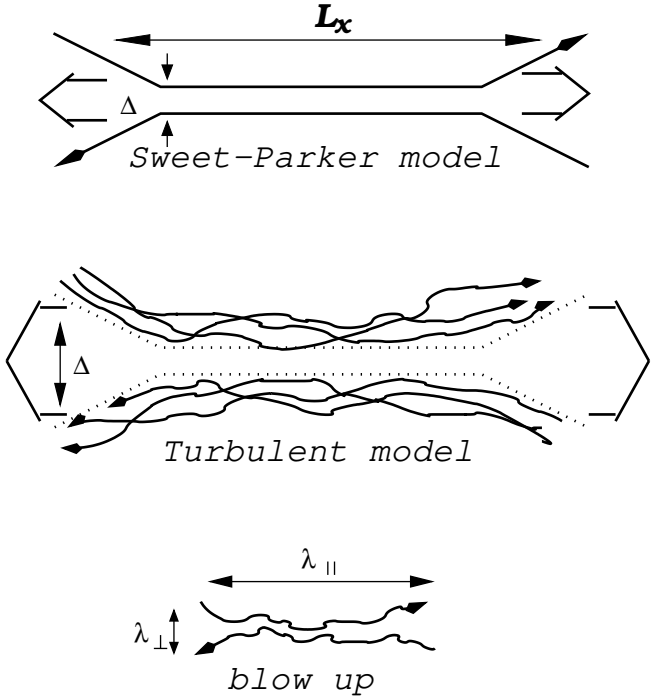}
\caption{{\it Upper plot}:
Sweet--Parker model of reconnection. The outflow
is limited by a thin slot $\Delta$ determined by Ohmic
diffusivity. The other scale is an astrophysical scale $L_{x} \gg \Delta$.
{\it Middle plot}: reconnection of weakly stochastic magnetic field according to
LV99. The outflow is limited by the magnetic field wandering.
{\it Low plot}: an individual small-scale reconnection region.
 From \citealt{lazarian_etal_2004}).}
\label{recon1}
 \end{center}
\end{figure}

The magnetic diffusion mechanism  that we address is the process deeply rooted in microscopic
 physics of how magnetic fields behave in highly conductive flows. The textbook
way to characterize these flows is to use the Lundquist number $S=L_x V_A/\eta$, where
$\eta$ is Ohmic diffusivity and $L_x$ and $V_A$ are the scale of system at hand and the Alfv\'en velocity, respectively. For astrophysical systems $L_x$ is very large and therefore $S\gg 1$. The brute force numerical study of such systems is not feasible as the
corresponding Lundquist numbers of the numerical simulations are much smaller.
However, if the Lundquist number of the flow does not control magnetic reconnection
this opens prospects of modelling magnetic diffusion in astrophysical systems.

The classical magnetic reconnection that follows the textbook Sweet-Parker scenario (see Figure 1, upper panel), depends on the Lundquist number and it is slow for astrophysical systems. Indeed, in the model all the matter moving with the speed $V_{rec}$ over the scale $L_x$ should be ejected with the Alfv\'en velocity through a thin slot $\Delta$. The disparity of astrophysical typical  scales $L_x$ and the scale
$\Delta$ determined by microphysics, i.e. resistivity, makes Sweet-Parker reconnection rate negligibly small for most of astrophysical applications, including the case of accretion disks. However, this scenario is
not valid in the presence of ubiquitous astrophysical turbulence (see Figure 1, lower panels). For the turbulence case, LV99 showed that the reconnection becomes independent of the resistivity, i.e. becomes  fast, as the outflow region $\Delta$ gets determined by
magnetic field wandering. This challenges the well-rooted concept of magnetic field frozenness for the case of turbulent fluids (see more in \citealt{eyink_etal_2011}, henceforth ELV11) and provides an interesting way of removing magnetic flux out of, e.g. accretion disks. The LV99 model was successfully tested numerically in \citet{kowal_etal_2009}.

The justification of the reconnection diffusion concept in L05 is based on
the LV99 model (see also \citealt{lazarian_etal_2004} for the case of magnetic
reconnection in  partially ionized gas).
Numerical effects are always a concern when dealing with reconnection and magnetic
field diffusion. Indeed, unlike the numerical tests in \citet{kowal_etal_2009}, in our simulations
the reconnection events are happening on small scales where numerical effects are
important. Precisely because of that, the numerical experiments with anomalous
 resistivity in \citet{kowal_etal_2009} are of key importance. There, using a numerical setup with
 high resolution in a magnetic reconnection layer, \citet{kowal_etal_2009} showed that in the presence of
turbulence the local non-linear enhancements of resistivity were not important.
This confirmed the corresponding analytical prediction in LV99 (see more discussion
in ELV11). Appealing to that finding, we claim that the reconnection diffusion that
 we observe in our simulations is a real effect and not a numerical
 artefact\footnote{As it was discussed in LV99 and shown even more explicitly in
 ELV11, the plasma effects that can enhance the local reconnection speed are not
important in the presence of turbulence which induces magnetic field wandering.}.
Analytical studies summarized in ELV11 also support the notion that magnetic
fields are generically not frozen-in when conductive fluids are turbulent. In
view of them, the concept of reconnection diffusion in L05 looks very natural.

As emphasized before, the concept above of magnetic flux transport in turbulent flows by reconnection diffusion has
been already successfully tested numerically by \citet{santos-lima_etal_2010} for idealized models of star
forming clouds. In the present paper, we study whether the same concept can entail
substantial changes for the magnetic field removal in the formation of 
protostellar disks. An extension to accretion disks in general can be also foreseen (see below).

\section{Numerical Setup and Initial Disk Conditions}

To investigate  the formation of a rotationally supported disk due
to turbulent reconnection magnetic flux transport, we have
integrated numerically  the following system of  MHD equations:

\begin{equation}
\frac{\partial \rho}{\partial t}  + \nabla \cdot \left( \rho \mathbf{u} \right)  = 0
\end{equation}
\begin{equation}
\rho \left( \frac{\partial}{\partial t} + \mathbf{u} \cdot \nabla \right) \mathbf{u} = - c_{s}^{2} \nabla \rho + \frac{1}{4 \pi}(\nabla \times \mathbf{B}) \times \mathbf{B} - \rho \nabla \Psi + \mathbf{f}
\end{equation}
\begin{equation}
\frac{\partial \mathbf{A}}{\partial t} = \mathbf{u} \times \nabla \times \mathbf{A} - \eta _{Ohm} \nabla \times \nabla \times \mathbf{A}
\end{equation}
where $\rho$ is the density, $\mathbf{u}$ is the velocity, $\Psi$ is the gravitational potential generated by the protostar,
$\mathbf{B}$ is the magnetic field, and $\mathbf{A}$ is the vector potential
 with $\mathbf{B = \nabla \times A + B}_{ext}$ (where  $\mathbf{B}_{ext}$
is the initial uniform magnetic field).
$\mathbf{f}$ is a random
force term responsible for the injection of turbulence. An isothermal
equation of state is assumed with  uniform sound speed $c_{s}$.

We solved the MHD equations above in a three-dimensional domain using a second-order
shock-capturing Godunov scheme  and  second-order Runge-Kutta  time
integration.  We employed a modified version of the code originally developed by \citet{kowal_etal_2007}, using the  HLL Riemann solver to
obtain the numerical fluxes  in each time step.

In order to compare our results with those of \citet{krasnopolsky_etal_2010}, we have considered the same initial conditions as in their setup.

Our  code works with cartesian coordinates and vector field
components. We started the system with a
collapsing cloud progenitor with initial constant rotation  (see below) and uniform
magnetic field in the $z$ direction.

 Given the cylindrical symmetry of the problem, we adopted  circular boundary conditions. Eight rows of ghost cells were put outside a inscribed circle in the $xy$ plane. For the four outer rows of ghost cells, we adopted fixed boundary conditions in the radial direction in every time step, while for the four inner ghost cells, linear interpolation   between the initial conditions and the values in the interior bound of the domain were applied for the density,  velocity, and vector potential. With this  implementation the  vector potential $\mathbf{A}$ has kept its initial null value. Although this produces some spurious noisy components of $\mathbf{B}$ in the azimuthal and vertical directions in the boundaries, these are too far from the central regions of the domain to affect the disk evolution. In the $z$ direction, we have applied the usual open boundary conditions (i.e., zero derivatives for all conservative quantities: density, momentum and potential vector). We found this implementation far more stable than using open boundary conditions in the $x$ and $y$ directions, or even  in the radial direction. Besides,  adopting  $circular$ rather than square boundaries prevented the formation of artificial spiral arms and $corners$ in the disk.
 
 
For modelling the accretion in the central zone, the technique of sink particles was
implemented in the code in the same way as described in \citet{federrath_etal_2010}. A central sink with accretion radius  encompassing 4 cells was introduced in the domain.
The gravitational
force inside this zone has a smoothing spline function identical to that presented in \citet{federrath_etal_2010}. We do not allow the
creation of sink particles elsewhere, since we are not calculating the
self-gravity of the gas.  We note that this accreting zone essentially provides a $pseudo$ inner boundary for the system and for this reason 
the dynamical equations are not directly solved  there where accretion occurs, although we assure momentum and mass conservation.

The physical length scales of the computational domain are
$6000$ AU in the $x$ and $y$ directions and $4000$ AU in the $z$
direction. A sink particle of mass $0.5$ M$_{\odot}$ is put in the center
of the domain. At  $t=0$, the gas has a uniform density
$\rho_{0} = 1.4 \times 10^{-19}$ g cm$^{-3}$ and a sound speed of $c_s =
2.0 \times 10^{4}$ cm s$^{-1}$ (which implies a temperature $T \approx 4.8  \overline{\mu}$ K, where $\overline{\mu}$ is mean molecular weight in atomic units). The initial rotation profile is
$v_{\Phi} = c_s \tanh (R/R_c)$ (as in \citealt{krasnopolsky_etal_2010}), where $R$ is the radial distance to
the central $z$-axis, and the characteristic distance $R_c = 200$ AU.

 We employed a uniform resolution of $384$x$384$x$256$ which for the chosen set of
 parameters  implies that each cell has a physical size of $15.6$ AU in each direction. The sink zone has an accretion  radius of $62.5$ AU. \footnote{We note that in the two-dimensional simulations of \citet{krasnopolsky_etal_2010}, they use a non-uniform mesh with a maximum resolution of $0.2$ AU in the central region. The employment of a non-uniform mesh could be advantageous in this problem, allowing a better resolution close the protostar. However, in the present work since we are dealing with turbulence injection in the evolving system, the use of a uniform mesh has the advantage of making the effects of numerical dissipation  more uniform and therefore, the analysis of the turbulent evolution and behavior   more straightforward.}

Although we are interested in the disk that forms inside a radius of approximately $400$ AU around  the central axis, we have carried out the simulations in a much larger region of $6000$ AU in order to keep the dynamically important central regions  of the domain free from any  outer boundary effects.

\section{Results}

We performed simulations for four  models which are listed in Table 1. Model \textit{hydro} is a purely hydrodynamical rotating system. All the other models have the same initial (vertical) magnetic field with intensity $B_z=35 \: \mu$G.  In order to have a benchmark, in the model named \textit{resistive} we
included  an $anomalous$ high  resistivity, with a magnitude about 3 orders of magnitude larger than the  Ohmic resistivity estimated for the system, i.e.,
 $\eta=1.2 \times 10^{20}$ cm$^{2}$ s$^{-1}$.
According to the results of \citet{krasnopolsky_etal_2010}, this is nearly the ideal value that the magnetic  resistivity  should  have in order to remove the magnetic flux excess of a typical collapsing  protostar disk progenitor and allow the formation of a rotationally sustained disk. We have thus included this anomalous resistive model in order to compare with more realistic MHD models that do not appeal to this resistivity excess.

We have also considered an MHD model with turbulence injection (labeled as  \textit{turbulent} model in Table 1). In this case, we introduced in the cloud  progenitor a solenoidal turbulent velocity field  with a characteristic scale of $1600$ AU and a Mach number $M_{S} \approx 4-5$ increasing approximately linearly from $t=0$ until $t=3 \times 10^{10}$ s (or $\approx 3000$ yr). These parameters result an estimated turbulent diffusivity which is of the order of the anomalous diffusivity employed in \textit{resistive} model: $\eta_{turb} \sim V_{turb} L_{inj} \sim 10^{20}$ cm$^{2}$ s$^{-1}$.
The induced turbulent velocity field has been intentionally smoothed  beyond a radius of $800$ AU,
 by a factor
$\exp \{ -[R(AU) - 800] ^2 / 400^2 \}$, in order to prevent disruption of the cloud at large
radii. The  injection of turbulence was stopped  at $t=4.5 \times 10^{11}$ s $\approx 0.015$ Myr. From this  time on,  it naturally decayed with time, as one should expect to happen in a real system when the physical agent that injects turbulent in the cloud ceases to occur.

The last of the models (which is labeled as ideal MHD) has no explicit resistivity or turbulence injected so that in this case the disk evolves under an ideal MHD condition.

\begin{table}[!hbt]
\caption{Summary of the models}
\centering
\begin{tabular}{l c c c c c}
\hline \hline
   &   &   $\eta_{Ohm}$   &   $\eta_{turb}$ \\
Model   &   $B_0$ ($\mu$G)   &   (cm$^{2}$ s$^{-1}$)   &   (cm$^{2}$ s$^{-1}$) \\
[0.5ex]
\hline
\textit{hydro}  &   $0$   &   $0$   &   $0$ \\
\textit{resistive} &   $35$   &   $1.2 \times 10^{20}$   &   $0$ \\
\textit{turbulent} &   $35$   &   $0$   &   $\sim 10^{20}$ \\
\textit{ideal MHD} &   $35$   &   $0$   &   $0$ \\
[1ex]
\hline
\end{tabular}
\end{table}

Figure 2 shows
face-on and edge-on density maps of the central slice of the disk for the four models at
 $\approx 0.03$ Myr.
The arrows in the top panels represent the direction of the velocity field, while those in the bottom panels represent the direction of the magnetic field.

The pure hydrodynamical model in the left panels  of Figure 2 clearly shows the formation of a high density torus structure within a radius  $\approx$ 300 AU which is typical of a Keplerian supported disk (see also Figure 3, top-right panel).

In the case of the ideal-MHD model (second row panels in Figure 2), the disk core is much smaller and a thin, low density outer  part extends to the outskirts of the computational domain. The radial velocity component is much larger than in  the pure hydrodynamical model.
The bending of the disk in the core region is due to the action of the magnetic torques. As the poloidal  field lines are dragged to  this region by the collapsing fluid,  large magnetic forces develop and act on the rotating flow. The resulting torque  removes angular  momentum from the inner disk and destroys its rotational support (see also Figure 3, upper panels).

The third column (from left) of panels  in Figure 2 shows the resistive MHD model. As in Model 1,
a torus (of radius $\approx 250$ AU) with a rotationally dominant velocity field is formed and  is surrounded by  a flat,
low density disk up to a radius of $\sim$ 500 AU.
Compared to  the ideal MHD model (second column), the structure of the magnetic field is much simpler and exhibits  the familiar hourglass geometry.

\begin{figure*}[!hbt]
 \begin{center}
 \includegraphics[width=1.0 \textwidth]{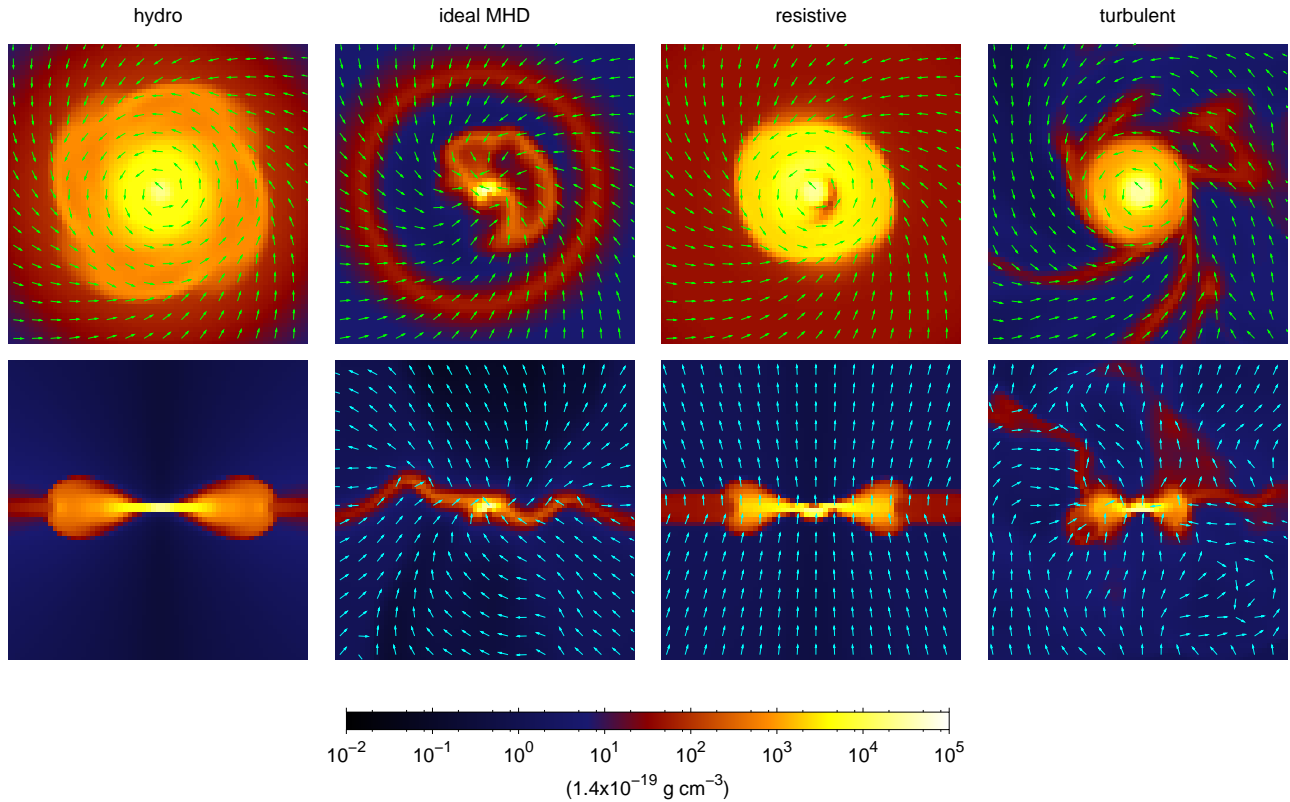}
 \caption{Face-on (top) and edge-on (bottom) density maps of the central slices of the collapsing disk models listed in Table 1 at a time $t=9 \times 10^{11}$ s ($\approx 0.03$ Myr).  The arrows in the top panels represent the velocity field direction an those in the bottom panels represent the magnetic field direction.  From left to  right rows it is depicted: (1) the pure hydrodynamic rotating system; (2) the ideal MHD model; (3) the MHD model with an anomalous resistivity $10^3$  times larger than the Ohmic resistivity, i.e. $\eta=1.2 \times 10^{20}$ cm$^{2}$ s$^{-1}$; and (4) the turbulent MHD model with turbulence  injected from $t=0$ until t=0.015 Myr. All the MHD models have an initial vertical magnetic field distribution with intensity $B_z=35 \: \mu$G. Each image has a side of $1000$ AU.}
 \end{center}
\end{figure*}


The last column (on the right  of Figure 2) shows  the ideal MHD model with injected turbulence (labeled \textit{turbulent}). A high density disk arises in the central region within a radius of 150 AU  surrounded by turbulent debris.
 From the simple visual inspection of the velocity field inside the disk  one cannot say if it is rotationally supported. On the other hand, the distorted structure  of the magnetic field in this region, which is  rather distinct from the helical  structure of the ideal MHD model, is an indication that  magnetic flux is being removed by the turbulence in this case. The examination of the velocity and magnetic field intensity profiles in Figure 3 are more elucidative, as described below.

Figure 3 shows  radial profiles of: (i) the radial velocity $v_R$ (top left), (ii)the rotational velocity $v_{\Phi}$ (top right); (iii) the inner disk mass (bottom left); and (iv) the vertical magnetic field $B_z$ (bottom right) for the models of Figure 2.
$v_R$ and $v_{\Phi}$ were averaged inside cylinders centered in the protostar
 with height $h=400$ AU and thickness $dr=20$ AU. Only cells with a density larger than $100$ times the initial density of the cloud ($\rho_{0}=1.4 \times 10^{-19}$ g cm$^{-3}$) were taken into account in the average evaluation.
 The internal disk mass was calculated in a similar way, but instead of averaging, we simply summed the masses of the cells in the inner region. The magnetic field profiles were also obtained from average values inside equatorial rings centered in the protostar with radial thickness $dr=20$ AU.


\begin{figure*}[!hbt]
 \begin{center}
 \includegraphics[width=1.0 \textwidth]{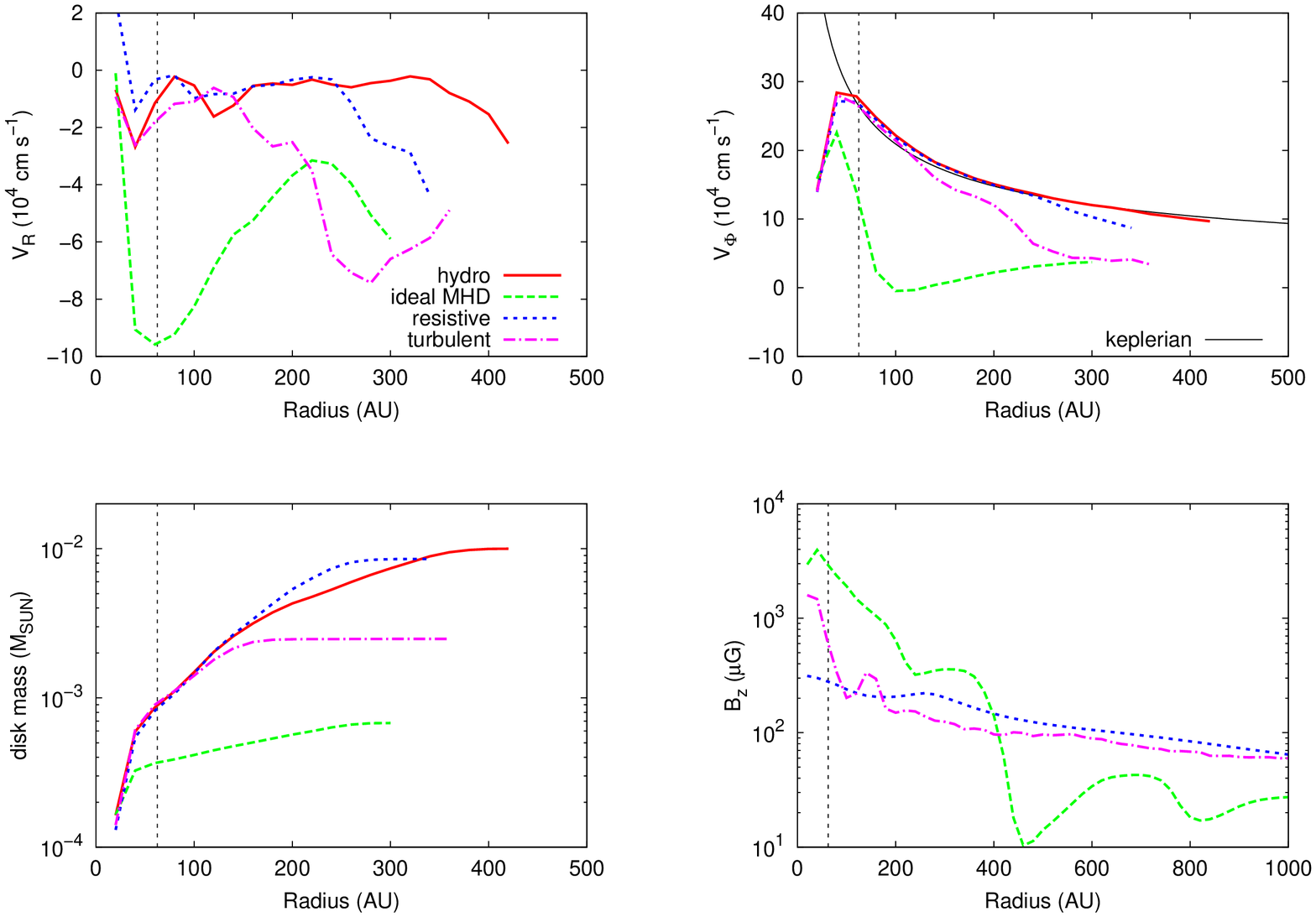}
 \caption{Radial profiles of the: (i)   radial velocity $v_{R}$ (top left), (ii)  rotational velocity $v_{\Phi}$ (top right); (iii) inner disk  mass (bottom left); and (iv)  vertical magnetic field $B_{z}$, for the four models of Figure 2 at time $t\approx 0.03$ Myr). The velocities were averaged inside cylinders centered in the protostar with height $h=400$ AU and thickness $dr=20$ AU.  The  magnetic field values were also  averaged  inside equatorial rings centered in the protostar. The standard deviation for the curves are not shown in order to make the visualization clearer, but they have typical values of: $2-4 \times 10^{4}$ cm s$^{-1}$ (for the radial velocity), $5-10 \times 10^{4}$ cm s$^{-1}$ (for the rotational velocity), and $100 \: \mu$G (for the magnetic field). The vertical lines indicate the radius of the sink accretion  zone.}
 \end{center}
\end{figure*}

For an ideal rotationally supported disk, the centrifugal barrier prevents the gas to fall  into the center. In this ideal scenario, the radial velocity should be null (at distances above the accretion sink zone). The top left panel of Figure 3 depicts the curves of the radial velocities  for the four models. Above the central sink accretion zone ($R> 62.5$ AU), the hydrodynamical model  (\textit{hydro}) is the prototype of a rotationally supported disk, the radial velocity being smaller than the sound speed ($c_S = 2 \times 10^{4}$ cm s$^{-1}$) inside the formed disk.
In the ideal MHD model,  the effect of the magnetic flux braking partially destroys the centrifugal barrier and the radial (infall)  velocity becomes very large, about three times the sound speed.  The MHD model with anomalous resistivity instead, shows a very similar behavior to the hydrodynamical model due to the efficient removal of magnetic flux from the central regions. In the case of the turbulent model, although it shows a persisting  non-null radial (infall) speed even above $R>62.5$ AU this is much smaller than in the ideal MHD model, of the order of the sound speed where the disk forms (between the R $\approx 70$ AU and $\approx 150$ AU).

The top right panel of Figure 3 compares the rotational velocities  $v_{\Phi}$  of the four models with the Keplerian profile $v_{K} = \sqrt{GM_{*}/R}$. All  models show similar trends to the Keplerian curve (beyond the accreting zone), except the ideal MHD model. In this case, the strong  suppression  of the rotational velocity due to  removal of angular momentum by the magnetic field to outside of the inner disk region is clearly seen, revealing a complete failure to form a rotationally supported disk. The turbulent MHD model, on the other hand, shows good agreement
 with the Keplerian curve at least inside the radius of $\approx 120$ AU where the disk forms.  Its rotation velocity profile is also very similar to the one of the MHD model with constant anomalous resistivity.
 Both models are able to  reduce the magnetic braking effects by removing magnetic flux from the inner  region  and the resulting rotation curves of the formed disks are nearly Keplerian. In the resistive model, this is provided by the hyper-resistivity, while in the turbulent model is the turbulent reconnection that provides this diffusion.



The bottom right panel of Figure 3 compares the profiles of the vertical component of the magnetic field, $B_z$, in the equator of the four models of Figure 2.
While in the ideal MHD model, the intensity and gradient of the magnetic field in the central regions are very large due to the inward advection  of magnetic flux by the collapsing material, in the anomalous resistive MHD model, the magnetic flux excess is completely removed from the central region resulting a smooth radial distribution of the field.
In the turbulent model, the smaller intensity of the magnetic field in the inner region and smoother distribution along the radial direction compared to the ideal MHD case are  clear evidences of the transport of magnetic flux to the  outskirts of the disk due to turbulent reconnection \citep{santos-lima_etal_2010}.
We note however that, due to the complex  structure which is still evolving, the standard deviation from the average value is very  large in the turbulent model, with a typical value of $100 \: \mu$G (and even larger for radii smaller than $100$ AU) which accounts for the turbulent component of the field.


Finally, the bottom left panel of Figure 3 shows the mass of the formed disks in the four models, as a function of the radius. In the hydrodynamical and the MHD resistive models (\textit{hydro} and \textit{resistive}), the mass increases until $R \approx 250$ AU and $\approx 350$ AU, respectively, and both have similar masses. The masses in the ideal MHD  and the turbulent MHD models (\textit{ideal MHD} and \textit{turbulent}) increase up to $R \approx 150$ AU and $\approx 250$ AU, respectively, and are smaller  than those of the other models. Nonetheless the turbulent MHD disk  has a total mass three times larger than that of the ideal MHD  model. 


\subsection{Discussion}

\subsection{Our approach and alternative ideas}

\citet{shu_etal_2006} (and references therein) mentioned the possibility  that the ambipolar diffusion can be substantially enhanced in circumstellar disks, but did not consider this as a viable solution. The subsequent paper of \citet{shu_etal_2007} refers to the anomalous resistivity and sketches the picture of
magnetic loops being reformed in the way of eventual removing magnetic flux. The latter process requires fast reconnection and we claim that in the presence of fast reconnection a more natural process associated with turbulence, i.e. magnetic reconnection diffusion can solve the problem.

\citet{krasnopolsky_etal_2010} and \citet{li_etal_2011} showed by means of  2D simulations that an effective
magnetic resistivity  $\eta  \gtrsim 10^{19}$ cm$^{2}$ s$^{-1}$ is needed for
neutralizing the magnetic braking and enable the formation of a
stable, rotationally supported, 100 AU-scale disk around a protostar.
The origin of this enhanced resistivity is completely unclear and the value above
is at least two to three  orders of magnitude larger than the estimated ohmic
diffusivity for these cores (e.g., \citealt{krasnopolsky_etal_2010}). On the other
hand, these same authors found that ambipolar diffusion, the  mechanism often
invoked to remove magnetic flux in star forming regions, is unable to provide
such required levels of diffusivity (see also \citealt{li_etal_2011}).

In this work, we have explored a different mechanism to remove the magnetic flux
excess from the central regions of a  rotating magnetized collapsing core which is
based on  magnetic reconnection diffusion in a turbulent flow. Unlike the Ohmic resistivity enhancement, reconnection diffusion does not appeal to any hypothetical processes, but to the turbulence existing in the system and fast magnetic reconnection of turbulent magnetic fields. One of the consequences of fast reconnection is that, unlike resistivity, it conserves magnetic field helicity. This may be important for constructing self-consistent models of  disks.

In order to compare our turbulent MHD model with other rotating disk formation
models, we  also performed 3D simulations of a pure hydrodynamical, an ideal MHD
and a resistive MHD model with a hyper-resistivity coefficient $\eta \sim 10^{20}$ cm$^{2}$ s$^{-1}$ (Figure 2). The essential  features produced in these three models
are in  agreement with the 2D models of  \citet{krasnopolsky_etal_2010}, i.e.,  the ideal MHD model is unable to produce a rotationally supported disk due to the
magnetic flux excess that accumulates in the central regions, while the MHD model
with artificially enhanced resistivity produces a nearly-Keplerian disk with
dimension, mass, and radial and rotational velocities similar to the pure
hydrodynamical model.

The rotating disk formed out of our turbulent MHD model exhibits  rotation velocity
and vertical magnetic field distributions along the radial direction which are
similar to the resistive MHD model (Figure 3). These similarities indicate that
the turbulent magnetic reconnection is in fact acting to remove the magnetic flux
excess from the central regions, just like the ordinary enhanced resistivity does
in the resistive model. We note, however, that the disk formed out of the turbulent
model is slightly smaller and less massive than the one produced in the
hyper-resistive model. In our tests the later has a diameter $\sim 250$ AU,
while the disk formed in the turbulent model has a
diameter $\sim 120$ AU which is  compatible with the observations.

The effective resistivity associated to the MHD turbulence in the turbulent model is approximately given by $\eta_{turb} \sim V_{turb} L_{inj}$, where $V_{turb}$ is the turbulent rms velocity, and $L_{inj}$ is the scale of injection of the turbulence
 \footnote{ We note however, that this value may be somewhat larger in the presence of  the gravitational field  (see L05, SX10)}.
We have adjusted the values of $L_{inj}$ and $V_{turb}$ in the turbulent model employing turbulent dynamical times ($L_{inj}/V_{turb}$) large enough to ensure that the cloud would not be destroyed by the turbulence before forming the disk. We tested several values of  $\eta_{turb}$ and the one employed in the model presented in Figure 2 is of the same order of the magnetic diffusivity of the model with enhanced resistivity, i.e.,  $\eta_{turb} \sim V_{turb} L_{inj} \approx 10^{20}$ cm$^{2}$ s$^{-1}$.
Smaller values were insufficient to produce rotationally supported disks.
Nonetheless, further  systematic parametric  study should  be performed in the future.


As mentioned in \S1, \citet{machida_etal_2010, machida_etal_2011}  have also performed 3D MHD simulations of disk formation and     obtained  a rotationally supported disk solution when including only  Ohmic resistivity  (with a dependence on density and temperature obtained from the fitting of the resistivities computed in \citealt{nakano_etal_2002}). However, they had to evolve the system much longer, about four times longer than in our turbulent simulation, in order to obtain a rotationally supported disk of 100-AU scale.  In their simulations, a tiny rotationally supported disk forms in the beginning because the large Ohmic resistivity  that is present in the very high  density inner regions is able to dissipate the magnetic fields there.  Later, this disk grows to larger scales due to the depletion of the infalling envelope.   Their initial conditions with a  more  massive gas core (which has a central density nearly  ten times larger   than in our models) probably  helped the formation of the rotating massive disk (which is almost two orders of magnitude more massive than in our turbulent model).  The comparison of our results  with theirs indicate that even though  at late stages Ohmic, or more possibly ambipolar diffusion, can become dominant in the high density cold gas, the turbulent diffusion in the early stages of accretion is able to form a light and large rotationally supported disk very quickly, in only a few $10^4$ yr. 

Finally, we should remark that other mechanisms to remove or reduce the effects of the magnetic braking in the inner regions of  protostellar cores have been also investigated  in the literature recently.  \citet{hennebelle_ciardi_2009} verified that the magnetic braking efficiency may decrease significantly when the rotation axis of the core is misaligned with the  direction of the regular magnetic field. They claim that even for small angles of the order of $10-20^o$ there are significant differences with respect to the aligned case. Also, in a concomitant work to the present one,  \citet{krasnopolsky_etal_2011} have examined the Hall effect on  disk formation. They found that a Hall-induced magnetic torque can diffuse magnetic flux outward and generate a rotationally supported disk in the collapsing flow, even when the core is initially non-rotating, however the spun-up material remains too sub-Keplerian \citep{li_etal_2011}.

Of course, in the near future, these mechanisms must be tested along with the just proposed turbulent magnetic reconnection and even with ambipolar diffusion, in order to assess  the relative importance of each effect on disk formation and evolution.
Nonetheless,  since MHD turbulence is expected to be present in these magnetic cores (e.g., \citealt{ballesteros-paredes_maclow_2002, melioli_etal_2006, leao_etal_2009, santos-lima_etal_2010}, and references therein), turbulent reconnection arises as  a natural mechanism for removing  magnetic flux excess and allowing the formation of these disks.

\subsection{Present work and SX10}

In  recent numerical
study in SX10, we showed that magnetic reconnection in a turbulent cloud can efficiently
transport magnetic flux from the inner denser regions to the periphery of the cloud
thus enabling the cloud to collapse to form a star.

Here, also by means of
fully 3D MHD simulations, we have
investigated  the same mechanism acting in  a rotating collapsing cloud core  and
shown that the magnetic flux excess of the inner regions of the system can be
effectively removed allowing the formation of a rotationally sustained
protostellar disk.

Another empirical finding in
SX10 is that the efficiency of the magnetic field expulsion via reconnection
 diffusivity increases with the source  gravitational field. This is a natural
consequence of diffusion
in the presence of the gravitational field which pulls one component (gas) and does not act on
the other weightless component (magnetic field). In terms of the problem in hand,
this implies that more massive protostars can induce magnetic field segregation even
for weaker level of turbulence. We plan to explore numerically this issue in a
forthcoming work.

\subsection{Present result and bigger picture}

In this paper we showed that the concept of reconnection diffusion (L05)
successfully works in the formation of  protostellar disks.
Together with our earlier testing of magnetic field removal through
reconnection diffusion from collapsing clouds this paper supports a
considerable change of the paradigm of star formation. Indeed, in the presence of
reconnection diffusion, there is no necessity to appeal to ambipolar diffusion. The
latter may still be important in low ionization, low turbulence environments, but, in
any case, the domain of its applicability is seriously challenged.

The application of the reconnection diffusion concept to protostellar disk formation and, in a more general framework, to accretion disks in general,  is natural as
the disks are expected to be turbulent, enabling our appeal to LV99 model of fast
reconnection. An important accepted source of turbulence in accretion disks is the
well known magneto-rotational instability (MRI) 
(Chandrasekhar 1960, Balbus \& Hawley 1991)
\footnote{We note, however, that in the present study, we were in a highly magnetized disk regime, where the magneto-rotational instability is ineffective.}, but at earlier stages turbulence can be
induced by the hydrodynamical  motions associated with the disk  formation. Turbulence is ubiquitous in astrophysical environments as it follows
from theoretical considerations based on the high Reynolds numbers of astrophysical
flows and is strongly supported by studies of spectra of the interstellar electron density fluctuations (see \citealt{armstrong_etal_1995, chepurnov_lazarian_2010}) as well as of  HI
 (\citealt{lazarian_2009} for a review and references therein; \citealt{chepurnov_etal_2010})
and CO lines (see \citealt{padoan_etal_2009}). 
The application of the reconnection diffusion mechanism to already formed accretion disks will be investigated in detail elsewhere. It should be noted however that former studies of the injection of turbulence in accretion disks have shown that at this stage turbulence may be ineffective to magnetic flux diffusion outward \citep{rothstein_lovelace_2008}.

\section{Summary}

Appealing to the LV99 model of fast magnetic reconnection and inspired by
the successful demonstration of removal of magnetic field
through reconnection diffusion from numerical models of molecular
clounds in SX10 we have performed numerical simulations and demonstrated that:

1. The concept of reconnection diffusion is applicable to the formation of protostellar disks with radius $\sim 100$ AU. The extension of this concept to accretion disks is foreseen.

2. In the gravitational field, reconnection diffusion mitigates  magnetic breaking
allowing the formation of protostellar disks.

3. The removal of magnetic field through reconnection diffusion is fast enough
to explain observations without the necessity of appealing to enhanced fluid resistivity.

\acknowledgments
RSL and EMGDP acknowledge partial support from grants of the Brazilian Agencies FAPESP (2006/50654-3 and 2007/04551-0), and CNPq (306598/2009-4).  AL acknowledges the NSF grant AST 0808118 and the NSF-sponsored Center for Magnetic Self-Organization. Part of the work was done by AL while in Germany supported by Humboldt Award. The authors are also grateful to G. Kowal for his very useful comments on the numerical implementation of the problem.

%
%

\end{document}